\documentclass{PoS}

\PoS{PoS(LAT2005)158}

\title{The chiral transition in two-flavor QCD}
\ShortTitle{The chiral transition in two-flavor QCD}

\author{Massimo D'Elia\\

	Dipartimento di Fisica,\\
	Universit\`a di Genova and INFN Sezione di Genova,\\
	Via Dodecaneso 33, I-16146 Genova, Italy\\
        E-mail: \email{delia@ge.infn.it}}

\author{Adriano Di Giacomo\\

	Dipartimento di Fisica,\\
	Universit\`a di Pisa and INFN Sezione di Pisa,\\
	largo Pontecorvo 3, I-56127 Pisa, Italy\\
        E-mail: \email{digiaco@df.unipi.it}}

\author{\speaker{Claudio Pica}\\

	Dipartimento di Fisica,\\
	Universit\`a di Pisa and INFN Sezione di Pisa,\\
	largo Pontecorvo 3, I-56127 Pisa, Italy\\
        E-mail: \email{pica@df.unipi.it}}

\abstract{
QCD with $N_f$=2 is a specially interesting system to investigate the
chiral transition. The order of the transition has still not been
established. We report the results of an in-depth numerical
investigation performed with staggered fermions on lattices with
$L_t$=4 and $L_s$=12,16,20,24,32 and quark masses $am_q$ ranging from
0.01335 to 0.307036. Using finite-size techniques we compare the
scaling behavior of a number of thermodynamical susceptibilities with
the expectations of $O(4)$ and $O(2)$ universality classes. Clear
disagreement is observed. Indications of a first order transition are
found. Preliminary reports of this work were presented at past Lattice
conferences.}

\FullConference{XXIIIrd International Symposium on Lattice Field Theory\\

                 25-30 July 2005\\

                 Trinity College, Dublin, Ireland}

\newcommand{\be}{\begin{equation}}
\newcommand{\ee}{\end{equation}}
\newcommand{\pbp}{\langle \bar{\psi} \psi \rangle}

\begin{document}

\section{Introduction}

The phase transition in $N_f=2$ QCD ($m_u=m_d\equiv m$) can provide
fundamental insight into the mechanism of confinement\cite{TWOFLVS}.

In the limit $m\rightarrow\infty$ the system is quenched, the
deconfining transition is first order, $\langle L\rangle$ is the order
parameter and $Z_3$ the symmetry. However the coupling to quarks breaks
$Z_3$.

At $m\simeq 0$ the deconfining transition seems to coincide with the
chiral transition. A transition line exists at intermediate values of
$m$ which is defined by the maxima of a number of susceptibilities
(the specific heat $C_V$, the susceptibility $\chi_m$ of the chiral
condensate $\pbp$, the susceptibility of $\langle L\rangle$) which all
coincide whithin errors (see e.g. Ref.\cite{karsch1}).

At $m\simeq 0$ a renormalization group analysis\cite{wilcz1} indicates
that the chiral transition is first order for $N_f\geq 3$, and for
$N_f=2$ can be either first order or second order in the universality
class of $O(4)$. For the latter case the transition is a crossover at
$m\neq 0$, and in particular a tricritical point is expected in the
$T-\mu$ plane\cite{Stephanov} which could be observed in heavy ion
collisions.

Various groups have studied this problem with Wilson\cite{cp-pacs} or
staggered\cite{fuku1,colombia,karsch1,jlqcd,milc} fermions. No clear
sign of discontinuities is found at least for the lattice sizes used,
but no agreement either with the critical exponents of $O(4)$.

We investigate the issue using standard Kogut-Susskind fermions on
lattices $4\times L_s^3$ with $L_s=12,16,20,24,32$ and lattice quark
masses $am_q$ ranging from $0.01335$ to $0.307036$. In the present
work we give our final results using our whole dataset collected on
APEmille machines and a detailed analysis which is different in many
points from the previous ones present in the literature. Preliminary
reports of this work were presented at past conferences.

\section{Analysis of the critical behavior}

The order of a phase transition can be investigated by finite size
scaling analysis. Near the critical point, for a second or a weak
first order transition, the singular behavior is described by power
law divergences according to universal critical exponents. The
singular part of the free energy density ${\cal F}_s$ is a homogeneuos
function of the reduced temperature $\tau\equiv 1-T/T_c$, the lattice
quark mass $am_q$ and the size of the system $L_s$:
\be
{\cal F}_s (\tau, am_q, L_s ) = L_s^{-d} {\cal F}_s \left(\tau L_s^{y_t}, am_q L_s^{y_h} \right)\label{FREENRG}
\ee
There are two independent critical exponents: the thermal $y_t$ and
the magnetic one $y_h$. The other more familiar critical exponents are
given by ($d=3$ is the dimensionality of the system):
\be
\alpha=2-\frac{d}{y_t} \hspace{1.2cm}
\beta=\frac{d}{y_t}-\frac{y_h}{y_t} \hspace{1.2cm}
\gamma=2\frac{y_h}{y_t}-\frac{d}{y_t} \hspace{1.2cm}
\delta=-\frac{y_h}{d-y_h} \hspace{1.2cm} \nu=\frac{1}{y_t}  
\ee
The numerical values of the critical exponents of interest for this work are given
in Table~\ref{CRITEXP}.

\begin{table}[b!]
\center\small
\begin{tabular}{|c|c|c|c|c|c|c|c|}
\hline & $y_t$ & $y_h$ & $\nu$ & $\alpha$ & $\gamma$ & $\beta$ & $\delta$\\
\hline $O(4)$ & 1.336(25) & 2.487(3) & 0.748(14) & -0.24(6) & 1.479(94) & 0.3837(69) & 4.852(24)\\
\hline $O(2)$ & 1.496(20) & 2.485(3) & 0.668(9) & -0.005(7) & 1.317(38) & 0.3442(20) & 4.826(12)\\
\hline $MF$ & $3/2$ & $9/4$ & $2/3$ & 0 & 1 & 1/2 & 3\\
\hline $1^{st} Order$ & 3 & 3 & $1/3$ & 1 & 1 & 0 & $\infty$\\
\hline
\end{tabular}
\caption{Critical exponents of $O(4)$, $O(2)$, mean field and first order.}\label{CRITEXP}
\end{table}

According to Eq.~\ref{FREENRG} this problem involves two scaling
variables. We simplify the problem restricting the parameter space in
three different ways:
\begin{enumerate}
\item keep $am_q L_s^{y_h}$ fixed while varying $L_s$.
One must choose a fixed value for $y_h$ so that only one universality
class at a time can be tested. Note that by chance the numerical value
of $y_h$ is the same for $O(4)$ and $O(2)$, so in practice we can test
these two critical behaviors at once. The scaling law for ${\cal F}_s$
is given by: ${\cal F}_s = L_s^{-d} {\cal F}_s \left(\tau L_s^{y_t}
\right)$ where now $am_q$ is a function of $L_s$;
\item keep $\tau$ and $am_q$ fixed and take the limit
$L_s\rightarrow\infty$ (finite mass scaling,
Ref.~\cite{karsch1,jlqcd,milc}). This amounts to assume the existence
of one single scale ($am_q$) in the problem, and no divergence of the
correlation length at the transition: i.e. finite mass scaling and no
finite size scaling. Eq.~(\ref{FREENRG}) becomes: ${\cal F}_s =
am_q^{d/y_h}{\cal F}_s (\tau am_q^{-y_t/y_h} )$ where the size of the
system $L_s$ is not present;
\item the last possibility is to keep $\tau L_s^{y_t}$
fixed while taking $L_s$ much bigger than the pion correlation lenght:
$aL_s\gg 1/m_\pi$. This should work better than case 2. if the
correlation lenght becomes large and comparable to $L_s$. The free
energy scaling law in this case is: ${\cal F}_s = am_q^{d/y_h} {\cal
F}_s \left(\tau L_s^{y_t} \right)$.
\end{enumerate}
In cases 2. and 3. we are not forced to fix $y_h$ so we are free to
test all the critical behaviors of interest. The price for this is
that we were forced to make additional approximations.

Taking appropriate derivatives we find the relevant scaling laws for
the thermodynamic susceptibilities\footnote{See Ref.~\cite{TWOFLVS}
for the detailed definition of the susceptibilities on the
lattice.}. In this work we concentrate on two of them: the specific
heat $C_V$ and the chiral susceptibility $\chi_m$. Note that while the
first quantity is always relevant, the second one is a good quantity
only if $\pbp$ is the order parameter: that is certainly true for
$m=0$ but may not be for $m>0$. The expected behavior for the cases
listed above are:
\begin{enumerate}
\item $C_V-C_0=L_s^{\alpha/\nu}\phi_c ( \tau L_s^{1/\nu} )$ ;
$\chi_m-\chi_0=L_s^{\gamma/\nu}\phi_\chi ( \tau L_s^{1/\nu}
)$. i.e. the peaks scale as a power of $L_s$ and the widths as
$L_s^{-1/\nu}$;
\item $C_V-C_0=(am_q)^{-\alpha/\nu y_h}\phi_c ( \tau (am_q)^{-y_t/y_h} )$
; $\chi_m-\chi_0=(am_q)^{-\gamma/\nu y_h}\phi_\chi ( \tau
(am_q)^{-y_t/y_h} )$. i.e. the peaks scale as a power of $am_q$ and the
widths as $am_q^{y_t/y_h}$;
\item $C_V-C_0=(am_q)^{-\alpha/\nu y_h}\phi_c ( \tau L_s^{1/\nu} )$ ;
$\chi_m-\chi_0=(am_q)^{-\gamma/\nu y_h}\phi_\chi ( \tau L_s^{1/\nu}
)$. i.e. the peaks scale as a power of $am_q$ and the width as $L_s^{-1/\nu}$.
\end{enumerate}

\begin{table}[b!]
\center\small
\begin{tabular}{|c||c|c|c|c||c|c|c|c|}
\hline
& \multicolumn{4}{|c||}{Run1} & \multicolumn{4}{|c|}{Run2}  \\
\hline $L_s$ & 12 & 16 & 20 & 32 & 12 & 16 & 20 & 32  \\
\hline $am_q$ & 0.153518 & 0.075 & 0.04303 & 0.01335 & 0.307036 & 0.15 & 0.08606 & 0.0267 \\
\hline \# Traj. & 22500 & 87700 & 14520 & 14500 & 25000 & 131390 & 16100 & 15100 \\
\hline $aL_s \cdot m_\pi$ & 11.9 & 11.0 & 10.0 & 8.9 & 11.3 & 15.8 & 14.8 & 12.4  \\
\hline
\end{tabular}
\caption{Run parameters for our Monte Carlo simulations. A more
detailed list can be found in Ref.~\cite{TWOFLVS}.}\label{RUNPAR}
\end{table}

To test different scaling hypotheses we have performed a huge amount
of Monte Carlo simulations, divided into three groups (see
Table~\ref{RUNPAR}). The first two are called Run1 and Run2 and they
satisfy the requirements of case 1., namely we have fixed $am_q
L_s^{y_h}$ to two different values and $y_h$ appropriate for $O(4)$
and $O(2)$. The third group is formed by two other simulations with
the following parameters: $L_s=16$, $am_q=0.01335$; $L_s=24$,
$am_q=0.04444$; with $10000$ trajectories each. Simulations were made
using the hybrid R algorithm sweeping a wide range of values of
$\beta$ for each value of $am_q$. The datasets collected were analysed
using the multi-histogram reweighting procedure.

\section{Results}

\begin{figure}[t!]
\center
\includegraphics*[width=0.45\textwidth]{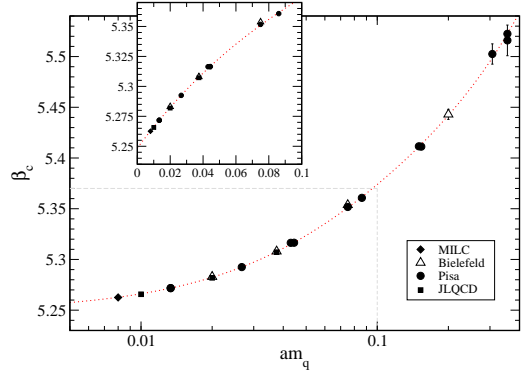}
\caption{Pseudocritical coupling values as a function of
$am_q$. The fit (explained in the text) has $\chi^2/dof=0.9$,
$dof=15$. Values obtained from other collaborations are also
included.}\label{CC}
\end{figure}
On the lattice the reduced temperature is expressed as a function of
the parameters $\beta$, $am_q$:
\be
\tau \equiv 1 - \frac{T}{T_0} =1 - \frac{a(\beta_0,0)}{a(\beta,am_q)}
\ee
where $a(\beta, am_q)$ is the lattice spacing. Expanding $a$ in power
series near $(\beta_0,0)$ we obtain:
\be
\tau \propto (\beta_0 - \beta )+ k_m am_q + k_{m^2} (am_q)^2 + k_{m\beta} am_q (\beta_0 - \beta ) + \cdots\label{TAU}
\ee
In previous works in the literature it was assumed that $\tau
\propto(\beta_0 - \beta )$ thus neglecting the lattice spacing
dependence on the lattice quark mass $am_q$. The terms in
Eq.~(\ref{TAU}) prove to be sufficient to describe the data of our
simulations. There are two different way the pseudocritical
temperature could scale: $\tau_c = k_\tau L_s^{1/\nu}$ or $\tau_c =
k'_\tau am_q^{y_t/y_h}$.  The expansion parameters and $k_\tau$ or
$k'_\tau$ can be obtained by fitting numerical data. The results are
the following: there is no visible shift of $\tau_c$ varying $L_s$,
i.e. $k_\tau$ is always compatible with zero; data are equally well
described by all the critical behaviors expected (first order, $O(4)$,
$O(2)$, MF) if one includes in the expression of $\tau_c$ the
quadratic terms shown above; including only the linear terms the
statement doesn't change if we restrict the allowed range of $am_q$
(see Ref.~\cite{TWOFLVS}); taking $\tau_c\propto \beta_0
- \beta_c$ does not describe the data in any mass range at our
disposal for a second order while for a first order it is identical to
keep only the linear mass term in Eq.~(\ref{TAU}).

\begin{figure}[t!]
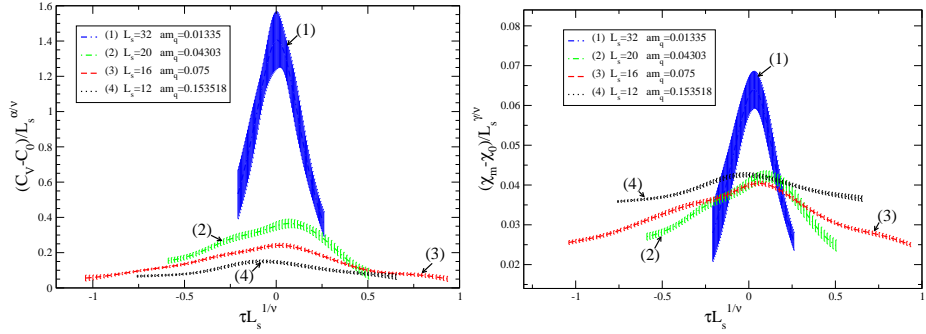

\center
\includegraphics*[width=0.4\textwidth]{Figs/Cv_Run1.eps}
\includegraphics*[width=0.4\textwidth]{Figs/Chi_Run1_2.eps}
\caption{Scaling behavior of $C_V$ and $\chi_m$ as expected in case
1. No sign of scaling is found. This excludes a second order $O(4)$ or
$O(2)$ transition. Simulations from Run1 are shown; Run2 is
similar\cite{TWOFLVS}.}\label{RUN1}
\end{figure}
\begin{figure}[b!]
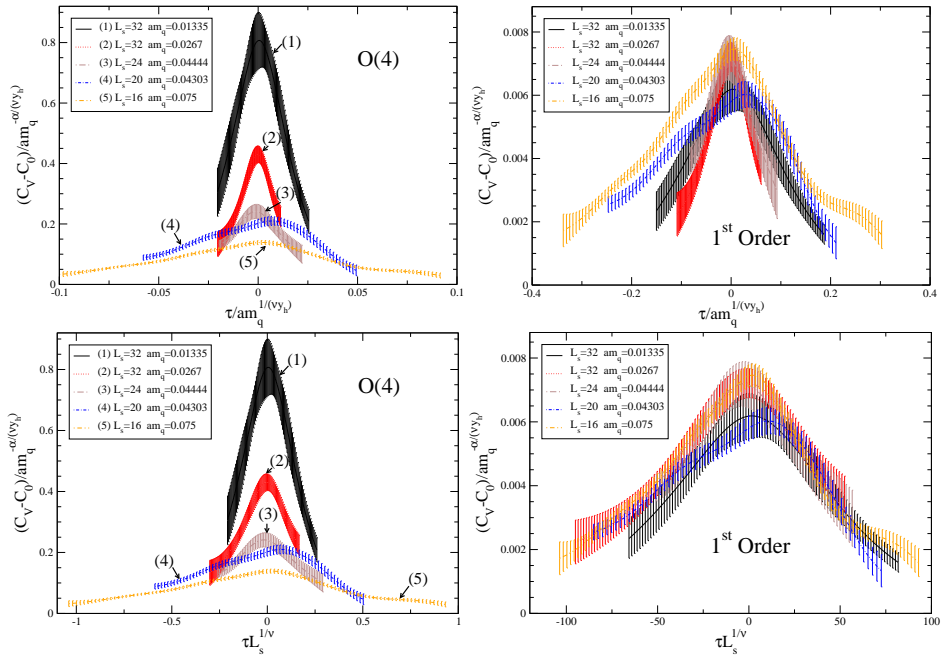

\center
\includegraphics*[width=0.4\textwidth]{Figs/Cv_O4.eps}
\includegraphics*[width=0.41\textwidth]{Figs/Cv_1st-m.eps}\\
\includegraphics*[width=0.4\textwidth]{Figs/Cv_O4-Ls.eps}
\includegraphics*[width=0.415\textwidth]{Figs/Cv_1st.eps}
\caption{Scaling of $C_V$ according to case 2. (top) and
3. (bottom) for both $O(4)$ (left) and first order (right). Other
second order behaviors are similar to the one shown.}\label{CASE23}
\end{figure}
Having fixed the parameters for the reduced temperature $\tau$ (see
Fig.~\ref{CC}) we can next consider the scaling of $C_V$ and
$\chi_m$. We first consider case 1: using data from Run1 and Run2 we
can check $O(4)$ and $O(2)$ critical behaviors. If there is scaling
all the curves obtained by plotting the susceptibilities divided by
the appropriate power of $L_s$ as functions of the scaling variable
$\tau L_s^{y_t}$ should collapse onto a universal scaling function.
As shown in Fig.~\ref{RUN1} no such scaling is observed for $O(4)$
($O(2)$ is quite similar). This clearly excludes these universality
classes. To check other critical behaviors we consider the scaling of
cases 2. and 3. We show in Fig.~\ref{CASE23} the analysis of $C_V$:
similar considerations also apply to $\chi_m$. The second order
critical behavior does not describe our data. The growth of the peaks
is instead consistent with a first order. This is an indication for a
first order transition. The widths of the curves seem to be described
by the scaling of case 3 and not according to case 2. This means that
the dependence on $L_s$ cannot be neglected. We have also explicitely
observed consistent deviations from the scaling predicted in case
2. making simulations at two different volumes with the same $am_q$,
see Ref.~\cite{TWOFLVS}. In fact this is due to the presence of two
different physical scales in the problem.

\begin{figure}[t!]
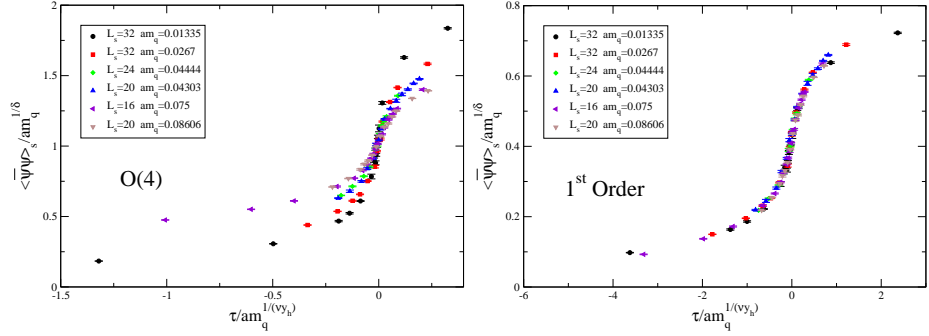

\center
\includegraphics*[width=.4\textwidth]{Figs/eqstO4-2.eps}
\includegraphics*[width=.4\textwidth]{Figs/eqst1st.eps}
\caption{Scaling of $\pbp$ for $O(4)$ (left) and first order (right).}\label{EQSTATE}
\end{figure}

One can also study the magnetic equation of state, i.e. the scaling of
$\langle\bar\psi\psi\rangle-\langle\bar\psi\psi\rangle_0 =
am_q^{1/\delta} F(\tau/am_q^{1/\nu y_h})$ as shown in
Fig.~\ref{EQSTATE}.
Also in this case the first order behavior seems to describe well the
data while the second order is excluded.

We have looked for metastabilities in the time histories as required
by a first order transition: no clear evidence exists up to the
lattice sizes explored.

\section{Conclusions and outlook}

We have investigated the nature of the chiral transition in $N_f=2$
QCD. By using the correct definition of $\tau$ on the lattice we have
shown that it is not possibile to discriminate the order of the
transition by looking only at $\beta_c$ with the present
data. Studying the critical behavior at fixed $am_q L_s^{y_h}$ we were
able to exclude the second order universality classes, i.e. $O(4)$,
$O(2)$, mean field.  We found evidence for a first order transition
looking at the scaling of thermodynamic susceptibilities but no clear
sign of discontinuities. The magnetic equation of state also seems
compatible with a first order transition. We plan to repeat the
analysis at fixed $am_q L_s^{y_h}$ for a first order transition
employing an improved action and $L_t=6$ to reduce possible artifacts.

\end{document}